\documentclass{spie}
            \usepackage{graphicx}
            \usepackage{amsmath,amssymb}
\begin{document}
            \title{Effects of mathematical locality and number scaling on  coordinate chart use}
            \author{Paul Benioff,\\
            Physics Division, Argonne National
            Laboratory,\\ Argonne, IL 60439, USA}
            \authorinfo{E-mail:pbenioff@anl.gov}

            \maketitle

            \begin{abstract}
             A stronger foundation for earlier work on the  effects of number scaling, and local mathematics  is described. Emphasis is placed on the effects of scaling on  coordinate systems. Effects of scaling are represented by a scalar field, $\theta,$ that appears in gauge theories as a spin zero boson. Gauge theory considerations led to the concept of local mathematics, as expressed through the use of universes, $\bigcup_{x},$ as collections of local mathematical systems at each point, $x,$ of a space time manifold, $M$. Both local and global coordinate charts are described.  These map $M$ into either local or global coordinate systems within a universe or between universes, respectively. The lifting of global expressions of nonlocal physical quantities, expressed by space and or time integrals or derivatives on $M$, to integrals or derivatives on  coordinate systems, is described.

             The assumption of local mathematics and universes makes  integrals and derivatives, on $M$ or on global charts, meaningless. They acquire meaning only when mapped into a local universe. The effect of scaling, by including the effect of $\theta$ into the local maps, is described. The lack of experimental evidence for $\theta$ so far shows that the coupling constant of $\theta$ to matter fields must be very small compared to the fine structure constant. Also the gradient of $\theta$ must be very small in the local region of cosmological space and time occupied by us as observers.  So far, there are no  known restrictions on $\theta$ or its gradient in regions of space and/or time that are far away from our local region.
            \end{abstract}


            \section{Introduction}
            The motivation for this work originates from Wigner's famous paper, "The unreasonable effectiveness of mathematics in the natural sciences" \cite{Wigner}, and comments about the ideas implied by the paper \cite{Omnes,Plotnitsky,Hamming}. This work has inspired others to investigate this problem from different angles \cite{Tegmark,Hut}. It also inspired attempts by this author to work towards a coherent theory of mathematics and physics together \cite{BenCTPM1,BenCTPM2}.

            A possible opening in this direction was the observation that in gauge theories, one starts with separate vector spaces at each point in space time \cite{Montvay}.  Matter fields, $\psi,$ take values, $\psi(x),$ in the vector space at location $x.$ The freedom of basis choice in the vector spaces, based on the work of Yang and Mills \cite{Yang}, led to the use of gauge group transformations to relate states in the different vector spaces. One result of this work was the development of the standard model, which has been so successful in physics.

            This work was extended by the observation that it is worth exploring consequences of replacing the one set of complex scalars, associated with all the vector spaces, by separate sets of complex scalars associated with each space time point. Extension of the freedom of basis choice in the vector spaces to include freedom of number scaling, led to the presence of a new spin $0$ scalar field, $\theta,$ in gauge theories  \cite{BenINTECH,BenNOVA,BenSPIE2}.

            Number scaling is based on the description of the different types of mathematical systems as structures \cite{Barwise,Keisler}. These consist of a base set of elements, and sets of basic operations, relations and constants that satisfy a relevant set of axioms. Number values can be scaled, provided the basic operations are scaled appropriately \cite{BenSCIRP}.

             The basic ideas behind the extension of gauge theories were  expanded by  first noting that other types of numbers, from the natural numbers to the real numbers, can be defined as substructures of the complex numbers. The fact that complex or other types of numbers are part of the description of many types of mathematical systems, vector spaces, algebras, groups, etc., led to the description of local collections or universes of mathematical systems at each point of a space and time manifold, $M$. Consequences of this assumption of local mathematics for theoretical descriptions of nonlocal physical properties were described.  The effect of number scaling on these properties and the resulting restrictions on the properties of the  scalar field, $\theta,$ were discussed  \cite{BenINTECH,BenNOVA,BenSPIE2}.

            This work expands the earlier work by paying more attention to the effects of coordinate charts that map $M$ onto space and time structures in the individual universes. The lifting of the description of nonlocal physical properties, such as  space and time integrals or derivatives based on $M,$ to those based on the space and time structures in the universes, is described.

            The work is organized by first giving, in section \ref{GTC}, a brief summary of the extension of gauge theories to include the effects of number scaling.  This is followed by a description of local universes, section \ref{LMU}. Included are descriptions of some general properties, of the universes, relations of observers to the universes, gauge theories in the universes, and maps between structures in different universes. Section \ref{STI} describes the effects of the presence of local universes on space and/or time integrals.  Both global and local descriptions are included.

            So far the effects of scaling on the integrals have not been included  in the local universes,  This is taken care of in the next section.  A final discussion section completes the paper.

            \section{Gauge theory considerations}\label{GTC}
            As noted, gauge theory considerations  were the original motivation for earlier work and for this work.   In gauge theories one starts with vector spaces, $\bar{V}_{x},$  at each point, $x$ of a space  time manifold, $M$ \cite{Montvay}.  Fields, $\psi,$ are maps from  $M,$ to points in the vector spaces where, for each point, $x,$ in $M,$ $\psi(x)$ is a vector in $\bar{V}_{x}.$

            Derivatives, $\partial_{x,\mu}\psi/\partial_{x}^{\mu}$, of the field at $x$ in the direction $\mu$, expressed as \begin{equation}\label{pdpsix}\frac{\partial_{x,\mu}\psi}{\partial_{x}^{\mu}} =\lim_{\Delta x^{\mu}\rightarrow 0} \frac{\psi(x+\Delta x^{\mu})-\psi(x)}{\Delta x^{\mu}},\end{equation} do not make sense.  The reason is that $\psi(x+\Delta x^{\mu})$ and $\psi(x)$ are in distinct vector spaces, $\bar{V}_{x+\Delta x^{\mu}}$ and $\bar{V}_{x}.$  Subtraction of vectors is not defined for vectors in different spaces.  It is defined only for vectors in the same space.

            This problem is  well known \cite{Montvay,Cheng}.  It is remedied by introducing a unitary gauge operator $U_{x,x+\Delta x^{\mu}}$ that maps $\psi(x+\Delta x^{\mu})$ in $\bar{V}_{x+\Delta x^{\mu}}$ into a vector, $U_{x,x+\Delta x^{\mu}}\psi(x+\Delta x^{\mu})$ in $\bar{V}_{x}$. The map also accounts for the freedom of basis choice of vectors in the neighboring vector spaces, an idea originally proposed by Yang Mills \cite{Yang} for isospin vector spaces and extended by  Utiyama \cite{Utiyama} to $n$ dimensional spaces.  The resulting covariant derivative, $D_{\mu,x}\psi,$ is given by\begin{equation} \label{Dmux}D_{\mu,x}\psi=\lim_{\Delta x^{\mu}\rightarrow 0} \frac{U_{x,x+\Delta x^{\mu}}\psi(x+\Delta x^{\mu})-\psi(x)}{\Delta x^{\mu}}.
            \end{equation}

            The earlier work \cite{BenINTECH,BenNOVA,BenSPIE2} begins with this description of covariant derivatives in gauge theories. It is noted that only one common complex, scalar field, $\bar{C}$ is associated with the different vector spaces $\bar{V}_{x}.$  This seems strange, especially since the basic axiomatic description of vector spaces includes scalar fields.  This suggested an extension of gauge theories by replacing the one common scalar field by local scalar fields, $\bar{C}_{x},$ associated with $\bar{V}_{x}$ for each $x$ in $M$.

            This change results in the expansion of the gauge group $U(n)$ for $n$ dimensional fields and vector spaces to $GL(1,R)\times U(n)$.  This adds an extra factor into the expression for the covariant derivative.  The operator $U_{x,x+\Delta x^{\mu}}$ in Eq. \ref{Dmux} is replaced by $R_{x,x+\Delta x^{\mu}}U_{x,x+\Delta x^{\mu}}.$ The factor, $R_{x,x+\Delta x^{\mu}},$ maps numbers in $\bar{C}_{x+\Delta x^{\mu}}$ to numbers in $\bar{C}_{x}$ just as $U_{x,x+\Delta x^{\mu}}$ maps vectors in $\bar{V}_{x+\Delta x^{\mu}}$ to vectors in $\bar{V}_{x}$. As will be noted shortly $R_{x,x+\Delta x^{\mu}}$ acts on vectors as well as on scalars.

             The usual use of gauge theories in physics and in the standard model already implicitly includes the operator $R_{x,x+\Delta x^{\mu}}$ for the special case that this operator is the identity operator. This means that  $R_{x,x+\Delta x^{\mu}}c(x+\Delta x^{\mu})$ is the same number value in $\bar{C}_{x}$ as $c(x+\Delta x^{\mu})$ is in $\bar{C}_{x+\Delta x^{\mu}}.$ For this case, replacement of the one set of complex numbers, $\bar{C}$ by local ones, $\bar{C}_{x}$ at each point has no effect.

             However this is a special case of a more general effect. Just as the gauge operator,  $U_{x,x+\Delta x^{\mu}}$ differs from the identity to account for the freedom of basis choice in the vector spaces, the operator, $R_{x,x+\Delta x^{\mu}}$ can differ from the identity to take account of the freedom of scaling of the complex numbers.

             \subsection{Number scaling}

             The possibility that numbers can be scaled is based on the mathematical logical definition of different types of mathematical systems as model structures \cite{Barwise,Keisler}. These structures consist of a base set of elements, a few basic operations, none or a few basic relations, and a few constants.  The properties of the structures must be such that the set of axioms, that are relevant to the system type being discussed,  must be true.  This description holds for all types of numbers, vector spaces, algebras groups, and many other system types.

             As examples, the structure for complex numbers is given by \begin{equation}\label{bC}
             \bar{C}=\{C,\pm,\times,\div,0,1\}.\end{equation} Here $C$ is the set of base elements as complex number values, and $0$ and $1$ are constants. The operations have their intended meaning. The structure, $\bar{C}$ must satisfy the axioms for an algebraically closed field of characteristic $0$ \cite{Complex}.

             The structure for (normed) vector  spaces, $\bar{V},$ is given by \begin{equation}\label{bV} \bar{V}=\{V,\pm,\cdot, ||-|| ,\psi\}\end{equation}Here $V$ is a base set of vectors, $\cdot$ denotes scalar vector multiplication, $||-||$ denotes the norm, and $\psi$ denotes an arbitrary vector.  it is not a constant. If the norm is derived from an inner product, $\langle -,-\rangle$ and the space is complete in the norm,  then $\bar{V}$ is a Hilbert space \cite{Kadison}.
             \begin{equation}\label{bH} \bar{H}=\{H,\pm,\cdot, \langle -,-\rangle,\psi\}\end{equation}  Here and in what follows, structures are distinguished from base sets by an overline.  Thus $\bar{C}$ is a structure with $C$ a base set.

             The possibility of number scaling is based on the observation that  complex number structures can be scaled by an arbitrary real number, $r\neq 0,$ and still preserve the truth of the relevant axioms. The scaled structure $\bar{C}^{r}$ is defined by
             \begin{equation}\label{bCr}\bar{C}^{r}=\{C_{r},+_{r},-_{r},\times_{r},\div_{r},0_{r},1_{r}\}=\{rC,+,-,\frac{\times}{r},r\div, 0,r\}.\end{equation} The first representation of $\bar{C}^{r}$ differs from $\bar{C}$  in Eq. \ref{bC}  by the presence of the subscript $r$ on each of the components.

              The subscript, $r,$ indicates that the components of $\bar{C}^{r}$ are scaled by a factor, $r,$ relative to the unscaled components of $\bar{C}.$ Details of the scaling are shown by the second representation of $\bar{C}^{r}.$ This representation shows the components of $\bar{C}^{r}$  in terms of the unscaled components in $\bar{C}.$ For example, multiplication in $\bar{C}^{r}$ corresponds to multiplication in $\bar{C}$  divided by $r,$  and division in $\bar{C}^{r}$ corresponds to division in $\bar{C}$ times $r.$ Also the number values that are zero and the identity  in the first representation of  $\bar{C}^{r}$ correspond to the number values $0$ and $r$ in the second representation.  Addition and subtraction are unaffected by scaling.

             The choice of the scaling factors  for the basic operations in $\bar{C}^{r}$ is not arbitrary.  They are chosen so that $\bar{C}^{r}$ satisfies the complex number axioms if and only if $\bar{C}$ does.  The proof that the definition of $\bar{C}^{r}$ satisfies this requirement is straight forward and will not be repeated here. As an example, the proof that $r$ is the multiplicative identity in $\bar{C}^{r}$ if and only if $1$ is the multiplicative identity in $\bar{C}$ is given by the following equivalences:\begin{equation}\label{Ideq}1_{r}\times_{r}c_{r}=c_{r}\Leftrightarrow r\frac{\times}{r}rc=rc \Leftrightarrow 1\times c=c.\end{equation}The first and second equations refer to the first and second representations of $\bar{C}^{r}$ and the third refers to $\bar{C}.$

             The invariance of the truth of the relevant axioms under scaling is an important requirement. It follows from this that complex analysis based on scaled complex numbers is equivalent to that based on unscaled numbers. Equivalence means that every unscaled element in analysis has a corresponding scaled element in the scaled number analysis, and conversely. Also the truth of equations in scaled analysis is preserved under mapping to the corresponding equations in unscaled analysis. For example, for any analytic function, $f,$ one has \begin{equation}\label{frcr}f_{r}(c_{r})=d_{r}\Leftrightarrow rf(c)=rd\Leftrightarrow f(c)=d. \end{equation}As was the case for Eq. \ref{Ideq}, the first and second equations refer to the first and second representations of $\bar{C}^{r},$ and the third one refers to $\bar{C}.$ These equivalences follow from the definition of analytic functions in terms of convergent power series \cite{BenSCIRP,Rudin}. The same description also holds for real analysis as real numbers are scaled in a similar fashion.

             Complex number scaling has an effect on vector spaces in general and on Hilbert spaces as a specific example.  The components of a scaled vector  space structure, $\bar{V}^{r}$ are given by  \begin{equation}\label{bVr}\bar{V}^{r}=\{V_{r}, +_{r},-_{r}, \cdot_{r},||-||_{r},\psi_{r}\}=\{rV,+,-,\frac{\cdot}{r},||-||,r\psi\}.\end{equation} Here the norm, in $\bar{V}$ of a vector, $r\psi$ with $r>0,$  is assumed to satisfy, $||r\psi||=r||\psi||.$ The righthand representation of $\bar{V}^{r}$ shows the representation of the components of $\bar{V}^{r}$ in terms of those of $\bar{V}.$ The corresponding representations of complex number structures, $\bar{C}^{r}$, and $\bar{C},$ are associated with the two representations of $\bar{V}^{r}.$

             For Hilbert spaces the scaled structures are given by \begin{equation}\label{bHr}\bar{H}^{r}=\{H_{r}, +_{r},-_{r}, \cdot_{r},\langle\psi_{r},\phi_{r}\rangle_{r},\psi_{r}\}=\{rH,+,-,\frac{\cdot}{r},\frac{\langle r\psi,r\phi\rangle}{r},r\psi\}.\end{equation} The righthand representation of $\bar{H}^{r}$ shows the representation of components of the scaled structure in terms of those of the unscaled $\bar{H}.$ As was the case for complex numbers, the scaled structures satisfy the relevant axioms if and only if the unscaled structures do.\footnote{There is another representation of $\bar{H}^{r}$ that limits scaling to just the complex numbers.  For this representation the right hand structure in Eq. \ref{bHr} is replaced by $\{H,\pm,\frac{\cdot}{r},r\langle\psi,\phi\rangle,\psi\}.$ This scaled representation is not used here because, for $n$ (and infinite) dimensional Hilbert spaces, this representation does not satisfy the well known equivalence, $\bar{H}^{r}\cong (\bar{C}^{r})^{n}.$ It is assumed here that this equivalence also extends to normed vector spaces.}

               The scaling of vector spaces, Eq. \ref{bVr}, affects the covariant derivative in Eq. \ref{Dmux}. This follows from the replacement of the vector spaces $\bar{V}_{x}$ by scaled vector spaces $\bar{V}^{r_{x}}_{x}$ where $r_{x}$ is a real positive space time dependent scaling factor.  In this case the action of the operator, $R_{x,x+\Delta x^{\mu}}$ becomes multiplication by the positive real number, $r_{x,x+\Delta x^{\mu}}=r_{x+\Delta x^{\mu}}/r_{x}.$

              The  space time dependence of the scaling factor can be represented by a scalar field, $\theta(x),$ as \begin{equation}\label{rxth}r_{x}=e^{\theta(x)}.\end{equation} In this case the factor, $R_{x,x+\Delta x^{\mu}}$  in the covariant derivative becomes multiplication by $e^{A_{\mu}(x)\Delta x^{\mu}}$.  Here the vector field, $\vec{A}(x),$ is the gradient of $\theta(x).$

              Note that the value of $\theta(x)$ has no effect on physics. The effect on physics is limited to the gradient of $\theta,$ or differences between the values of $\theta(x)$ at different points of space and time.

              Use of this extension of the covariant derivative  with the requirement that all Lagrangians be invariant under local $U(n)$ gauge transformations, as is done in the usual development of gauge theories,  gives the result that the field $\theta$ is  a spin $0$ scalar boson. There are no restrictions on whether or not the boson has mass.

              The fact that, so far, one cannot make a definite association of $\theta(x)$ to any existing or proposed scalar field in physics,  places some limitations on the properties of the field.  One is that the ratio of the coupling constant for this field to fermion fields must be very small compared to the fine structure constant. This follows from the fact that there is no evidence for an effect of $\theta$ in the very accurate description of experiments by QED.

              Another  restriction on $\theta$ is that the variation of $\theta(x)$ in the region of cosmological space and time occupied by us, as intelligent observers, must be very small.  However, there are no restrictions on the space and time dependence of $\theta(x)$ at cosmological distances that are far away  in space and or time  from our local region.  Details are given in \cite{BenNOVA}. Nevertheless there are many possible candidates for the physical description of $\theta.$ These include the Higgs boson, the inflaton, quintessence, and  other scalar fields that are proposed in the literature on cosmology  \cite{Higgs,PLorenz,Linde}.

             \section{Local mathematical universes}\label{LMU}

             So far,  mathematical structures associated with each point, $x$ of $M$, have been described for the complex numbers and vector spaces.  This notion of local structures can be extended to many other types of mathematical systems. It includes the natural numbers, integers, rational numbers, and real numbers as these can all be described as substructures of the complex numbers. The local structures for these types of numbers are given explicitly by
             \begin{equation}\label{NTx}\begin{array}{l} \bar{N}_{x}=\{N_{x},+,\times_{x},<_{x}, 0_{x},1_{x}\}\\\bar{I}_{x}=\{I_{x},\pm_{x},\times_{x},<_{x}, 0_{x},1_{x}\}\\ \overline{Ra}_{x}=\{Ra_{x},\pm_{x},\times_{x},\div_{x},<_{x}, 0_{x},1_{x}\}\\\bar{R}_{x}=\{R_{x},\pm_{x},\times_{x},\div_{x},<_{x}, 0_{x},1_{x}\}\\  \bar{C}_{x}=\{C_{x},\pm_{x},\times_{x},\div_{x}, 0_{x},1_{x}\}\end{array}\end{equation}

              The concept of local structures can be extended  to many other system types. These include all types of systems that include numbers as scalars in their description.  Included are vector spaces, algebras, group representations, coordinate charts, and many other types of systems.

              As a generic example, the local representation of a type $S$ system at point $x$ is given by \begin{equation}\label{bSx} \bar{S}_{x}=\{S_{x},Op_{S,x},Rel_{S,x},K_{S,x}\}. \end{equation}   Here $S_{x}$ is a base set of elements, $Op_{S,x},Rel_{S,x}$ and $K_{S,x}$ denote sets of basic operations, relations and constants for type $S$ systems. This structure must also satisfy a set of axioms for type $S$ systems. If the description of type $S$ systems involves other system types,  as is the case for vector spaces, then local structures for these systems, with  their relevant axioms, must be included with the description of $\bar{S}_{x}.$

              These and other local mathematical structures can be collected together into local  mathematical universes at each point, $x,$  of a space and/or time manifold, $M$.  A local universe, $\bigcup_{x}$,  includes all types of systems whose model structures have local representations, as in Eqs. \ref{NTx} and \ref{bSx}. It also includes many other types of  structures. Besides vector spaces, algebras, and group representations, a local universe contains, quantum and classical fields. It also contains expressions for all physical quantities that can be expressed as functions,  integrals, or derivatives over local coordinate representations of $M$. The nonlocality of these examples occurs within $\bigcup_{x}$.

            Coordinate system representations of $M$ in $\bigcup_{x}$ are expressed by use of charts that map open subsets of $M$ onto open subsets of  $\overline{R^{4}}_{x}$ \cite{Charts}. Here $M$ is assumed to be such that, for each $x,$ there are single charts that map all of $M$ onto $\overline{R^{4}}_{x}$. Manifolds that require atlases of charts to cover all of $M$, as is the case for general relativity, are excluded here.

            The real number labels in a coordinate system, $\phi_{x}(M)$ on $\bar{R}^{4}_{x},$ are real numbers in $\bar{R}_{x}.$ Here $\phi_{x}$ denotes a chart at $x.$ Even though the domain of each $\phi_{x}$ is all of $M$, $\phi_{x}$ is referred to here as a local chart. The reason is that it is associated with a point, $x,$ of $M$ and the values of $\phi_{x}$ are locations in $\bar{R}^{r}_{x}.$

             For the purposes of this work it is not necessary to specify $M$ in detail.  It can either be the space time of special relativity, as shown in the representation, $\bar{R}^{4}_{x},$ of $M$ at $x$, or Euclidean space and/or time.

             As far as this work is concerned, $M$ is  or represents, the  physical space and/or time of the physical universe. It is the space and time in which material bodies move and interact. Observers, as intelligent beings, are included as moving and interacting material bodies.

             In the following, the variables, $x,y,$ with no subscripts, will always refer to locations in $M$. Variables with subscripts, such as $u_{x},v_{x},$ will always refer to locations on a coordinate system, $\phi_{x}(M),$ in $\bigcup_{x}.$

             The description of gauge theories, given in the last section, can be lifted into any local universe. In $\bigcup_{x},$ the space time, $M,$ as the domain of various entities, is replaced by a coordinate chart, $\phi_{x}(M),$ that maps $M$ onto $\bar{R}^{4}_{x}.$ For each location $u_{x}$  in $\phi_{x}(M),$ a field, $\psi_{x}$ has a value, $\psi_{x}(u_{x}),$ in a vector space $\bar{V}_{u_{x}}$ at point, $u_{x}.$ For $n$ dimensional vector spaces, the gauge transformations, appearing in the covariant derivatives, are elements of the Lie algebra, $su(n)_{x}.$ The scalar field, $\theta(x)$ that appears in the number scaling factor is replaced by $\theta_{x}(u_{x}).$ Additional aspects and properties of gauge theories are easily included into $\bigcup_{x}$ by making appropriate changes of locations from $M$ to those in $\phi_{x}(M),$ and adding  the subscript, $x,$ where needed.

             \subsection{Some general aspects of local universes}

             It is best to regard each local mathematical universe as an open collection of mathematical systems.   One reason is that specification of the totality of mathematical systems in each $\bigcup_{x}$ is not possible, nor is it even desirable. Local universes can be expanded by inclusion of new mathematical systems that are defined or discovered by mathematicians and physicists, and this process is expected to be unending.

              The openness  of the universes  represents a change from the view used in previous work \cite{BenINTECH,BenNOVA,BenSPIE2}.  There the universes were regarded as closed collections containing all the mathematics that an observer can know in principle. The arguments given above strongly suggest that these universes can never be closed.

             Even though the local universes are open and cannot be specified in detail, they do have some interesting properties. As noted above, they do contain structures for all types of numbers and for many types of mathematical systems that include numbers in their description. Also all structures in $\bigcup_{x}$ have in common the fact that they are local  on $M$. They are all associated with point $x$ of $M$.

             Universes at different locations are equivalent in the sense that for each mathematical system type $S,$ there are structures,  $\bar{S}_{y}$ in $\bigcup_{y}$ and $\bar{S}_{x}$ in $\bigcup_{x}$ that are equivalent.  This is expressed by means of an isomorphic map $F_{S,x,y}$ that maps $\bar{S}_{y}$ onto $\bar{S}_{x}.$ If $s_{y}$ is an element in $\bar{S}_{y}$ then $s_{x}=F_{S,x,y}(s_{y})$ has the same value in $\bar{S}_{x}$ as $s_{y}$ does in $\bar{S}_{y}.$ Also $F_{S,x,y}(Op_{S,y})=Op_{S,x}$ and $F_{S,x,y}(Rel_{S,y})=Rel_{S,x}.$

             The presence of local universes and local structures places a restriction on the concept of equations that say that two terms are equal.  In all of the following, all valid equations relate elements that are within the same structure and within the same universe. The few invalid equations that do not satisfy this requirement will be described as meaningless or not defined.

             \subsection{Observers and local universes}

             There is an aspect of the relation between observers and mathematics that supports the idea of local mathematical universes. Mathematical logic emphasizes the distinction between syntactic and semantic structures in mathematics.  Semantic structures and their relations,  as models of relevant axiom sets, are the ones that are described in various branches of mathematics and theoretical physics.  As such they are required to have meaning.

             The "meaning of 'meaning'" in mathematics is a subject of much discussion \cite{MM}. Here there is no intention to delve into this topic.  The only aspect of meaning that is relevant is that statements and properties of mathematical (and physical) systems are meaningful only in the minds of observers as intelligent beings. In the absence of observers, no meaning can be given to mathematical (or physical) statements or to their truth or falseness.

             The brain of an observer is the seat of meaningfulness in that statements and equations are meaningful to an observer only when registered in the brain.\footnote{Details on how they are registered are of great interest.  However this is outside the purview of this work.}  Mathematical information in textbooks or spoken by a lecturer is meaningless until it has entered the the observers brains.

             It follows from this, and the relatively small size of a brain, that meaningfulness is a local concept limited to the size of the brain.  Here, to make things simple, it is assumed that observers and their brains occupy points in the physical space and time manifold, $M$. As a result, the mathematics that can have meaning to an observer, $O_{x}$ at point $x$ of $M,$ is limited to the structures and systems at $x$. This is the mathematics in $\bigcup_{x}.$ If the observer moves on a world path, $p(\tau),$ then at proper time $\tau$ the collection of systems that can have meaning to $O_{p(\tau)}$ are those in $\bigcup_{p(\tau)}.$

             The use of the expression, "can have meaning" instead of "have meaning" shows that this is a one way implication only. The implication, 'If structure $\bar{S}_{x}$ has meaning to  $O_{x},$ then it is in $\bigcup_{x}$', is valid. The converse is not necessarily valid because an observer may not understand, or know very little  mathematics.

\begin{figure}[h!]\begin{center}\vspace{2cm}
            \rotatebox{270}{\resizebox{170pt}{170pt}{\includegraphics[170pt,200pt]
            [520pt,550pt]{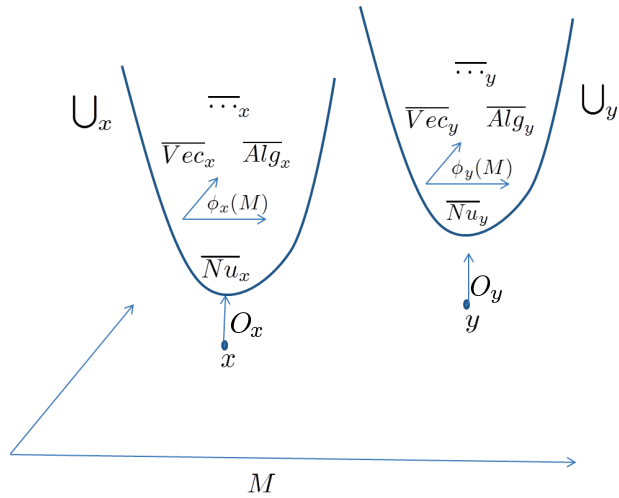}}}\end{center}\caption{Schematic illustration of different mathematical universes at different points in a Manifold $M$.  For ease of illustration the manifold is assumed to be two dimensional.  Two examples of the universes at points $x$ and $y$ are shown. In each universe, $\phi(M)$  with subscripts,  denotes the coordinate systems, also shown schematically, $\overline{Nu}$ denotes, collectively, structures for  the different types of numbers, and $\overline{Vec}$  and $\overline{Alg}$ denote, collectively, structures for the different types of vector spaces and algebras. The dots with overline denote the presence of many other structures. $O_{x}$ and $O_{y}$ denote observers at positions, $x,$ and $y.$}\label{SPIE41}\end{figure}

             The relations of observers to  the underlying manifold, $M,$ and to local universes can be made clear by means of a figure, shown here as figure \ref{SPIE41}. This figure shows  two universes and two observers at points $x$ and $y$ for a $2$ dimensional flat manifold, $M.$ The two are representative of the infinite number of universes, each associated with a point of $M$. Coordinate systems are represented as are collective representations of  a few of the different types of systems. The overlined dots stand for many other types of mathematical systems in the universes.

            It should be noted that these local universes  are assumed to exist at all points, $x$ of $M,$ independent of the presence of an observer.  The fact that properties of mathematical structures and the mathematics of  theoretical physics  in $\bigcup_{y}$  can have meaning to an observer at $x$ is accounted for by their equivalent representations in $\bigcup_{x}$.  For instance,  a structure $\bar{S}_{y}$ in $\bigcup_{y}$ is represented as the structure, $\bar{S}_{x,y}$ in $\bigcup_{x}.$ Here $\bar{S}_{x,y}$ is the view of $\bar{S}_{y}$ in $\bigcup_{x},$ as seen by $O_{x}.$ This is clearly necessary if $\bar{S}_{y}$ is to be registered in the brain of $O_{x}.$ The structure, $\bar{S}_{x,y},$ is equivalent to $\bar{S}_{x}$ in the sense that the value of each element, $s_{x,y}$  in $\bar{S}_{x,y}$ is the same as is the value of $s_{x}$ in $\bar{S}_{x}.$ also the operations and relations in $\bar{S}_{x,y}$ are the same as those in $\bar{S}_{x}.$

            In this case,  $\bar{S}_{x,y}$ can be identified with $\bar{S}_{x}$ and they can both be regarded as the same structure.  However, as will be seen later, for many structures, this is a special case of a more general one in which $\bar{S}_{x,y}$ is not equivalent to $\bar{S}_{x}.$

            The same description holds for the view that  $O_{y}$ has of structures in $\bigcup_{x}.$ The structure, $\bar{S}_{y,x}$ is the representation of $\bar{S}_{x}$ in $\bigcup_{y}.$ It is equivalent  to $\bar{S}_{y}$ and can be identified with $\bar{S}_{y}.$ As noted above, this is a special case of a more general representation.

             \subsection{Gauge theories in local universes}\label{GTLU}

             The description of  gauge theories in Section \ref{GTC} was based on fields and covariant derivatives that are based on points of $M$. This description can be lifted into each local universe by replacing $M$ by a coordinate system, $\phi_{x}(M),$ in $\bigcup_{x}.$  In this case, the covariant derivative, $D_{\mu.x}\psi$ in Eq. \ref{Dmux} has a local representation in $\bigcup_{x}$ as \begin{equation} \label{Dmuyx}(D_{\mu,u_{x}})_{x}\psi_{x}=\lim_{\Delta u_{x}^{\mu}\rightarrow 0} \frac{(U_{u_{x},u_{x}+\Delta u_{x}^{\mu}})_{x}\psi_{x}(u_{x}+\Delta u_{x}^{\mu})-\psi_{x}(u_{x})}{\Delta u^{\mu}_{x}}.\end{equation} The integration variable, $x,$ is replaced by the variable $u_{x}$ in $\phi_{x}(M)$.  The subscript $x$ is retained to indicate membership in $\bigcup_{x}.$

             This change of integration variable and interpretation of $x$ applies to the other expressions and results of gauge theories. The vector space $(\bar{V}_{x})_{u_{x}},$ has a local complex scalar field, $(\bar{C}_{x})_{u_{x}}$ associated with it.  Here local means local with respect to points on $\phi_{x}(M)$ in $\bigcup_{x}.$

             Number scaling is introduced  by replacing $(U_{u_{x},u_{x}+\Delta u_{x}^{\mu}})_{x}$ in Eq. \ref{Dmuyx} by $$e^{(A_{x}^{\mu}(u_{x})\Delta u^{\mu}_{x})}(U_{u_{x},u_{x}+\Delta u^{\mu}})_{x}.$$ Use of this  in gauge theory gives a result that is similar to that already obtained.  $\vec{A}_{x}$ is a vector field on $\phi_{x}(M)$ and is the gradient of a scalar field, $\theta_{x},$ which is a spin $0$ boson. For each $u_{x}$ in $\phi_{x}(M),$ the field, $\theta_{x},$ has values, $\theta_{x}(u_{x}),$ in $(\bar{R}_{x})_{u_{x}}.$

             \subsection{Maps between structures in different universes}

             As was noted, the local universes are equivalent in that for any structure $\bar{S}_{y}$ in $\bigcup_{y},$ there is an equivalent structure, $\bar{S}_{x}$ in $\bigcup_{x}$ and conversely.  This holds for any pair, $x,y$ of points in $M$.

             For any structure type, $S,$ this equivalence can be expressed by  isomorphic maps. For structures, $\bar{S}_{y}$ at $y$ and  $\bar{S}_{x}$ at $x$, the map $F_{S,x,y}$ is defined from Eq. \ref{bSx} as\begin{equation}\label{FxybS} \begin{array}{l} F_{x,y}(\bar{S}_{y})=\{F_{x,y}(S_{y}),F_{x,y}(Op_{y}), F_{x,y}(Rel_{y}),F_{x,y}(K_{y})\}\\\\\hspace{1cm} =\{S_{x},Op_{x},Rel_{x},K_{x}\}=\bar{S}_{x}.\end{array}\end{equation}To save on notation, the subscript $S$ has been removed from the map designation.

             Depending on ones viewpoint, the map $F_{x,y}$ either corresponds to or defines the notion of 'same value as', 'same operation as', and 'same relation as'. For each element, $s_{y}$ in $S_{y}$, $F_{x,y}(s_{y})=s_{x}$ is the same element in $S_{x}$ as $s_{y}$ is in $S_{y}.$ For each operation, $op_{y}$ in $Op_{y},$ $F_{x,y}(op_{y})=op_{x}$ is the same operation in $Op_{x}$ as $op_{y}$ is in $Op_{y}$. The same equivalence holds for the relations and constants.

             If $\bar{S}_{y}$ is a structure that uses other structures in its description, then the map $F_{x,y}$ must be extended to include these other structures. As an example, the isomorphic map $F_{H,x,y}$ of the Hilbert space, $\bar{H}_{y}$ onto $\bar{H}_{x}$ (both of the same number of dimensions) includes a map $F_{C,x,y}$ of $\bar{C}_{y}$ onto $\bar{C}_{x}$ as part of its description. The map $F_{H,x,y}$ is defined from Eq. \ref{bH} by \begin{equation}\label{FxybHy}\begin{array}{l}F_{H,x,y}(\bar{H}_{y})= \{F_{H,x,y}(H_{y}),F_{H,x,y}(\pm_{y}),\\\\\hspace{1cm}F_{HC,x,y}(\cdot_{y}),F_{C,x,y}(\langle-,-\rangle_{y}), F_{H,x,y}(\psi_{y})\}=\bar{H}_{x}.\end{array}\end{equation} The map component, $F_{HC,x,y}(\cdot_{y})$ is defined by $$F_{HC,x,y}(c_{y}\cdot_{y}\alpha_{y})=F_{C,x,y}(c_{y})F_{HC,x,y}(\cdot_{y})F_{H,x,y}(\alpha_{y})=c_{x}\cdot_{x}\alpha_{x}.$$ Here $c_{x}$ and $\alpha_{x}$ are the same complex number in $\bar{C}_{x}$ and vector in $\bar{H}_{x}$ as are the number, $c_{y}$ in $\bar{C}_{y}$ and the vector, $\alpha_{y}$ in $\bar{H}_{y}.$ Also $$F_{C,x,y}(\langle\alpha_{y},\psi_{y}\rangle_{y})=\langle F_{H,x,y}(\alpha_{y}),F_{H,x,y}(\psi_{y})\rangle_{x}=\langle \alpha_{x},\psi_{x}\rangle_{x}$$ is the same complex number in $\bar{C}_{x}$ as $\langle\alpha_{y},\psi_{y}\rangle_{y}$ is in $\bar{C}_{y}.$

             \section{Space and/or time integrals}\label{STI}

             These maps are sufficient for mapping many expressions in physics from one universe to their equivalents in other universes.  They are especially useful for describing physical quantities that are non local.  These  are quantities that include derivatives and integrals over space and/or time in their description.

             So far derivatives have been discussed for gauge theories. Here attention is turned to integrals.  There are two types of integrals over space and or time.  Local ones over  coordinate systems in a universe, $x$, and nonlocal ones over the manifold, $M$.  Local ones are discussed first.

             \subsection{Local space and/or  time integrals}

             A simple example of a local space and/or time integral is that of a complex valued field, If $f_{y}$ is a complex valued field  in $\bigcup_{y}$ such that, for each location, $u_{y}$, in $\phi_{y}(M),$ $f_{y}(u_{y})$ is a complex number in $\bar{C}_{y},$ then the integral of $f_{y}$ over $\phi_{y}(M)$ is given by \begin{equation}\label{Ify}I_{f,y}=\int f_{y}(u_{y})du_{y}.\end{equation}

             The corresponding integral over $\phi_{x}(M)$ of the corresponding function in $\bigcup_{x}$ is given by \begin{equation}\label{Ifx}I_{f,x}=\int  f_{x}(u_{x})du_{x}.\end{equation}The requirement that $I_{f,x}$ is the same integral in $\bigcup_{x}$ as is $I_{f,y}$ in $\bigcup_{y}$ includes  conditions that must be satisfied. These conditions are that the chart, $\phi_{y}$ and function $f_{y}$ must be the same in $\bigcup_{y}$  as are the chart $\phi_{x}$ and function, $f_{x}$ in $\bigcup_{x}.$ If this is the case, then the numerical value of $I_{f,y}$ in $\bigcup_{y}$ is the same as the value of $I_{f,x}$ in $\bigcup_{x}.$ This can be  expressed by\begin{equation}\label{FCIfxy}F_{I,x,y}(I_{f,y})=I_{f,x}.\end{equation}

             There are two conditions that must be satisfied so that $\phi_{x}(M)$ is  the same coordinate system in $\bigcup_{x}$ as $\phi_{y}(M)$  is in $\bigcup_{y}.$   One  is that the tuple, $u_{y}$ of real number labels in $\phi_{y}(M)$ is the same as is the tuple, $u_{x},$ in $\phi_{x}(M).$ This is expressed by use of an isomorphic map, $F_{R^{4},x,y}$ that takes $\overline{R^{4}}_{y}$ onto $\overline{R^{4}}_{x}$ such that for each quadruple, $u_{y}$ in $\overline{R^{4}}_{y},$
             \begin{equation}\label{FR4xy}F_{R^{4},x,y}(u_{y})=\{F_{R,x,y}(u_{y,\mu}):\mu=1,2,3,4\}=\{u_{x,\mu}:\mu=1,2,3,4\}.\end{equation}

              The other condition is that $u_{y}$ and $u_{x}$ must correspond to the same point on $M$.  This is satisfied by the requirement that, for each $u_{y}$ in $\phi_{y}(M),$ $\phi_{x}(\phi_{y}^{-1}(u_{y}))=u_{x}.$  These two conditions can be combined into the expression\begin{equation}\label{phixphiy} \phi_{x}(\phi_{y}^{-1}(u_{y})) =u_{x}=F_{R^{4},x,y}(u_{y}).\end{equation}

             The condition that $f_{y}$ is the same function as $f_{x}$ is given by the equation, \begin{equation}\label{FCxyfyu}
             F_{C,x,y}(f_{y}(u_{y}))=f_{x}(\phi_{x}\phi_{y}^{-1}(u_{y})))=f_{x}(u_{x}).\end{equation} This equation must hold for all pairs $u_{y},u_{x}$ of number tuples that satisfy Eq. \ref{phixphiy}.

\begin{figure}[h!]\begin{center}\vspace{2cm}
            \rotatebox{270}{\resizebox{170pt}{170pt}{\includegraphics[170pt,200pt]
            [520pt,550pt]{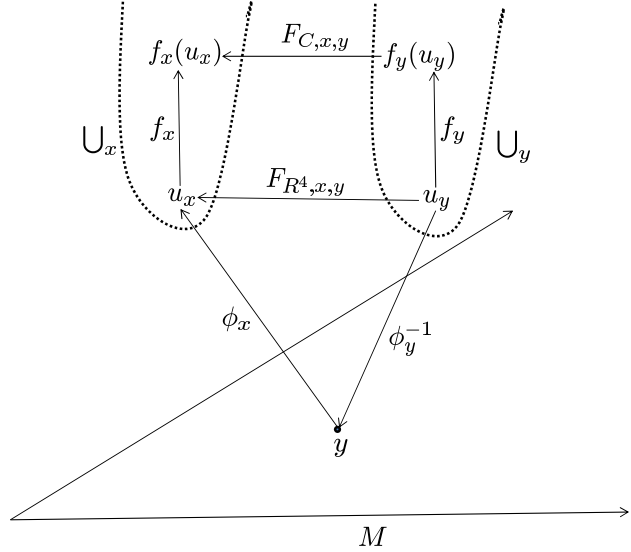}}}\end{center}\caption{Schematic illustration of the relations shown in Eqs. \ref{phixphiy} and \ref{FCxyfyu}. The underlying manifold is shown as two dimensional for ease in presentation. The universes are shown with dotted lines. The map $F_{R^{4},x,y}$ is for the full $4$ dimensional space time. The point, $x,$ which is the location of $\bigcup_{x}$, is not shown. }\label{SPIE42}\end{figure}

             Since the relations described in Eqs. \ref{phixphiy} and \ref{FCxyfyu} may be confusing, they and their relations to the containing universes are illustrated in Figure \ref{SPIE42}.

            Using these relations enables one to express $F_{I,x,y}$ as a composition of these relations as \begin{equation}\label{FIxyIfy}
            F_{I,x,y}I_{f,y}=\int_{x} F_{C,x,y}(f_{y}(u_{y}))F_{d,x,y}(du_{y})=I_{f,x}.
            \end{equation}Here $F_{d,x,y}(du_{y})=du_{x}$ is the same short distance in $\phi_{x}(M)$ as $du_{y}$ is in $\phi_{y}(M).$

             So  far, local descriptions of the space time integrals of the same  functions, $f_{x}$ at different locations, $x,$ on $M$ have been described. The existence of separate universes at each point of $M$ leads one to consider the possibility of describing a global integral of a global function, $f.$

             \subsection{Global space and/or time integrals}

             Let $\{\phi_{x}:x\epsilon M\}$ be a family of charts on $M$. The charts are all the same in that for each pair, $x,y$ of locations on $M$ $\phi_{x}$ and $\phi_{y}$ satisfy Eq. \ref{phixphiy}. Let $\{f_{x}:x\epsilon M\}$ be a family of complex valued functions where for each $x$ in $M$, $f_{x}$ is a function in $\bigcup_{x}$ with domain $\phi_{x}(M)$ and range  in or on $\bar{C}_{x}.$ The functions  in the family are the same in that for each pair, $x,y$ of points on $M$ $f_{x}$ is the same as $f_{y}$ in that they  satisfy Eq. \ref{FCxyfyu}.

             Define a global chart type map, $\phi,$ from the family of charts by \begin{equation}\label{phiphiyy}\phi(y)=\phi_{y}(y). \end{equation} The map, $\phi$ is not strictly a chart because the values of $\phi$ are in different local coordinate systems.  However it is equivalent to a chart because for any $x$ on $M,$ $F_{R^{4},x,y}(\phi(y))=\phi_{x}(y)$ for all $y$ in $M$.

             Define $f_{\phi}$ to be a function with domain, $\phi(M),$ and range such that for each $y$ in $M$,\begin{equation}\label{fphiy} f_{\phi}(\phi(y))=f_{y}(\phi(y))=f_{y}(\phi_{y}(y)).\end{equation} Here $f_{y}$ is a function in the family defined earlier. Let $I_{f,\phi}$ be the integral of $f_{\phi}$ over $\phi(M).$ It is given by  \begin{equation}\label{Ifphi}I_{f,\phi}=\int f_{\phi}(\phi(y))d\phi(y)=\int f_{y}(\phi(y))d\phi(y).\end{equation} This  integral is meaningless because $f_{\phi}(\phi(y))$ and $f_{\phi}(\phi(x))$ are complex numbers in different structures. Addition of these numbers, implied by the definition of integrals, is not defined as  addition is defined only within a structure, not between structures. Also  $d\phi(y)$ is undefined.

              The fact that $d\phi(y)$ is not defined follows from noting that \begin{equation}\label{dphiy}d\phi(y)=\phi(y+dy)-\phi(y) =\phi_{y+dy}(y+dy)-\phi_{y}(y). \end{equation}  The righthand subtraction is undefined because $\phi_{y+dy}(y+dy)$ and $\phi_{y}(y)$ are locations in coordinate systems at different sites on $M.$ As was the case for the values of $f,$ subtraction of coordinate values is not defined between coordinate systems at different locations.  It is defined only within a coordinate system at a given location.

             These problems with $I_{f,\phi}$ can be fixed by choosing a reference location, $x,$  mapping the values of the integrand to the same values in $\bigcup_{x}$ and then doing the integration. The first step is to map $d\phi(y)$ to $d\phi_{x}(y)$ This is achieved by noting that \begin{equation}\label{dphixy}F_{R^{4},x,y+dy} (\phi(y+dy))-F_{R^{4},x,y}(\phi(y))=\phi_{x}(y+dy)-\phi_{x}(y)=d\phi_{x}(y).\end{equation} This follows from the definitions of $F_{R^{4},x,y},$ Eq. \ref{FR4xy}, and $\phi,$ Eq. \ref{phiphiyy}.

             Use of  $F_{C,x,y}$  and $F_{R^{4},x,y}$ to  map  $f_{y}(\phi(y))$ to the same value, \begin{equation}\label{FCxyfy}
             F_{C,x,y}(f_{y}(\phi(y)))=f_{x}(F_{R^{4},x,y}(\phi(y)))=f_{x}(\phi_{x}(y)),\end{equation}
             in $\bigcup_{x}$ as $f_{y}(\phi(y))$ is in $\bigcup_{y},$ gives the result that \begin{equation}\begin{array}{l}\label{FIxphi} F_{I,f,x,\phi}(I_{f,\phi})=\int_{x}F_{C,x,y}(f_{y}(\phi(y)))F_{R^{4},x,y,y+dy}(d\phi(y))\\\\\hspace{1cm}=\int_{x}f_{x}(F_{R^{4},x,y} (\phi(y)))d\phi_{x}(y)=\int_{x} f_{x}(u_{x})du_{x}=I_{f,x}.\end{array}\end{equation} Here  $F_{R^{4},x,y,y+dy}$ is defined from equation \ref{dphixy} by $$d\phi_{x}(y)=F_{R^{4},x,y,y+dy}(d\phi(y)).$$

\begin{figure}[h!]\begin{center}\vspace{2cm}
            \rotatebox{270}{\resizebox{170pt}{170pt}{\includegraphics[190pt,200pt]
            [540pt,550pt]{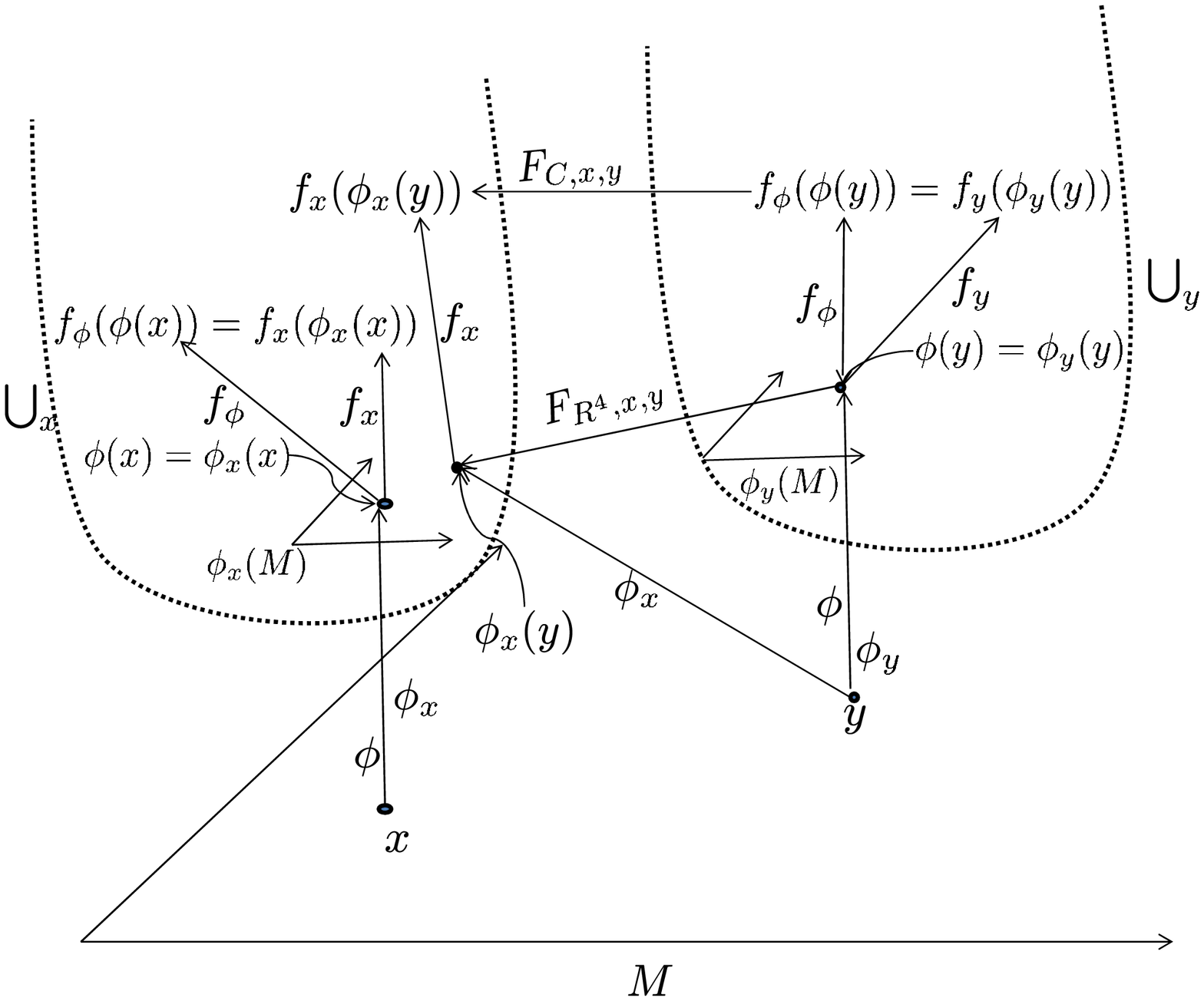}}}\end{center}\caption{Relations between the global and local functions and maps used in the integrals. $f_{\phi}$ and $\phi$ are global functions and coordinate charts that take values in the complex number structures and coordinate systems at different locations in $M$, This is shown in detail for two points, $x$ and $y.$ $f_{x},$ $f_{y},$ and $\phi_{x},$ $\phi_{y}$ are local functions and local coordinate charts that take values in $\bar{C}_{x},$ $\bar{C}_{y}$ and in $\phi_{x}(M)$, $\phi_{y}(M)$ respectively. The universes of systems at $x$ and $y$ are also shown with dotted lines.} \label{SPIE43}\end{figure}

             Figure \ref{SPIE43} shows the relations between the various parameters for these global functions just as Fig. \ref{SPIE42} does for the local expressions.  The figure has a lot of material on it.  However it should be an aid to understanding the  results.

             This result shows that, for each $x$ in $M$, the global expression, $I_{f,\phi}$ of the  integral of $f$ over $\phi(M)$ can be given meaning by mapping the integrand to $\bigcup_{x}$ and then integrating.  The result obtained is identical to the local integral of $f_{x}$ over $\phi_{x}(M),$ as given in Eq. \ref{Ifx}.

             This equivalence between maps of global space and/or  time integrals to local  integrals is expected to hold for many physical quantities. For example, consider the local expression at $x$ for a quantum wave packet, $\psi_{x}$ as
             \begin{equation}\label{psix}\psi_{x}=\int |u_{x}\rangle_{x}\psi_{x}(u_{x}) du_{x}.\end{equation} The integral is over all locations, $u_{x}$ in $\phi_{x}(M).$  Here $M$  is a three dimensional Euclidean space. The integrand, $h_{x}=|-\rangle_{x}\psi_{x}(-),$ can be regarded as a Hilbert space valued function, $h_{x},$ that takes values, $h_{x}(u_{x})$ in $\bar{H}_{x},\bar{C}_{x}$ for each $u_{x}$ in $\phi_{x}(M).$ For each $x,$ $\bar{H}_{x}$ is a Hilbert space in $\bigcup_{x}$ that can be used to describe wave packets as integrals over $\phi_{x}(M).$

             One can proceed as was done for the complex valued function $f.$  Define a family, $\{h_{x}:x\epsilon M\}$ of vector valued functions that are pairwise the same. This means that if $u_{y}$ and $u_{x}$ are related by Eq. \ref{phixphiy}, then $h_{y}(u_{y})$ is the same vector in $\bar{HC}_{y}\equiv \bar{H}_{y},\bar{C}_{y}$ as $h_{x}(u_{x})$ is in $\bar{HC}_{x}.$ This is expressed by the use of isomorphic maps as\begin{equation}\label{FHxyh}F_{HC,x,y}(h_{y}(u_{y}))=h_{x}(F_{R^{3},x,y}(u_{y})) =h_{x}(u_{x}).\end{equation}

             Define a global function, $h,$  on $\phi(M)$ by $h(\phi(y))=h_{y}(\phi(y))=h_{y}(\phi_{y}(y)).$  For each $y,$ $h(\phi(y))$ is a vector in $\bar{HC}_{y}$ in $\bigcup_{y}.$ As was the case for $f,$ the global integral, \begin{equation}\label{psi}
             \psi_{\phi}=\int h(\phi(y))d\phi(y)=\int |\phi(y)\rangle_{y}\psi_{y}(\phi(y))d\phi(y)\end{equation} is not defined.  However it can be mapped to a local integral at $x$ that is the same as is the integral for $\psi_{x}$.

             The map is similar to that  given in Eq. \ref{FIxphi}.  One obtains \begin{equation}\begin{array}{l} \label{FIhxphi} F_{I,h,x,\phi}\int h(\phi(y))d\phi(y)=\int_{x} F_{HC,x,y}(h_{y}(\phi(y)))F_{R^{3},x,y,y+dy}(d\phi(y))\\\\\hspace{1cm} =\int_{x}h_{x}(F_{R^{3},x,y} (\phi(y)))d\phi_{x}(y)=\int_{x}h_{x}(\phi_{x}(y))d\phi_{x}(y)\\\\\hspace{1cm} =\int_{x} |u_{x}\rangle_{x} \psi_{x}(u_{x})du_{x} =\psi_{x}.\end{array}\end{equation}

             One sees that this example is similar to the one for the function $f$ in that the local representation, at $x,$ of the global integral of the global function, $h,$  over $\phi(M),$ is identical to the local integral of the local function, $h_{x},$ over $\phi_{x}(M).$

             \section{Effects of scaling  on integrals}\label{SEI}

             The results obtained so far show that local representations of nonlocal physical quantities, such as integrals over space and/or time, are the same as local representations of global descriptions of these physical quantities.  This equivalence does not  hold if scaling is present. This was already seen for derivatives in gauge theories.  Here the affect on integrals is outlined.

             Scaling is present whenever mathematical elements must be moved from one location to another on $M.$ This effect will clearly be present in integrals of quantities over $M$. The movement is needed because, as was seen, integrals are defined only locally within a  universe, $\bigcup_{x},$ at some point, $x.$

             A good way to describe the effects of scaling is by a factorization of the maps $F_{S,y,x}$ that map $\bar{S}_{x}$ isomorphically onto $\bar{S}_{y}$.  Note that $F_{S,y,x}$ is the inverse of $F_{S,x,y}.$ For the complex numbers,  $F_{C,y,x}$ factors according to \begin{equation}\label{FCyxbC}\bar{C}_{y}=F_{C,y,x}(\bar{C}_{x})=Z_{C,y,x}W_{C,y,x}\bar{C}_{x}=Z_{C,y,x}\bar{C}^{r}_{x}. \end{equation}Here \begin{equation}\label{bCrx}  \bar{C}^{r}_{x}=\{C^{r}_{x},\pm^{r}_{x},\times^{r}_{x},\div^{r}_{x},0^{r}_{x}, 1^{r}_{x}\}=\{rC_{x},\pm_{x}\frac{\times_{x}}{r},r\div_{x},0_{x},r_{x}\}\end{equation} is the scaled representation of $\bar{C}_{y}$ in $\bigcup_{x}.$ The right hand representation shows the components of $\bar{C}^{r}_{x}$ in terms of those of $\bar{C}_{x}.$

             The factor, \begin{equation}\label{rryx} r\equiv r_{y,x}=\frac{(r_{y})_{x}}{r_{x}}=e^{\theta(y)_{x}-\theta(x)}\end{equation} is the scaling factor, where $r_{x}$ and $(r_{y})_{x}$ are defined in Eq. \ref{rxth}. Here $(r_{y})_{x}$ is the same real number value in $\bar{R}_{x}$ as $r_{y}$ is in $\bar{R}_{y},$ and  $\theta$ is a scalar field defined on $M$.

             In what follows it is useful to  first lift $\theta$ to an equivalent scalar field, $\theta_{\phi},$ defined over the global coordinate system values in  $\phi(M).$ Also let $\{\theta_{y}:y\epsilon M\}$ be a family of local scalar fields that are all the same. For each $y$, $\theta_{y}$ is a real valued scalar field in $\bigcup_{y}$ with domain $\phi_{y}(M)$ and such that for each $y$ in $M$,
              $\theta(y)=\theta_{y}(\phi_{y}(y)).$     Then $\theta(y)\equiv\theta_{\phi}(\phi(y))=\theta_{y}(\phi_{y}(y)).$ The corresponding change in the definition of $r_{y,x}$ is given by  \begin{equation}\label{ryxphi} r_{y,x}\equiv r_{\phi(y),\phi(x)}= e^{\theta(\phi(y))_{x}-\theta(\phi(x))}.\end{equation} Here $\theta(\phi(y))_{x}$ is the same value in $\bar{R}_{x}$ as $\theta(\phi(y))$ is in $\bar{R}_{y}.$

              Let $f$ be the global function, as defined in Eq. \ref{fphiy}, with domain $\phi(M).$ For each $u$ in $\phi(M),$ $f(u)$ is a complex number in $\bar{C}_{y}.$   Here $y$ is the location in $M$ such that $\phi(y)=u.$ Equivalently, $y=\phi^{-1}(u).$

              Let $u$ and $v$ be coordinate values defined by $\phi(y)=u$ and $\phi(x)=v.$ Addition of a complex number, $f(u),$ in $\bar{C}_{y}$ to $f(v)$ in $\bar{C}_{x},$ with scaling included, corresponds to addition, at $x,$ of the scaled representation, $f(u)^{r}_{x},$ of $f(u)$ at $x,$ to $f(v).$   $f(u)^{r}_{x}$ is related to $f(u)_{x}$ by  \begin{equation}\label{fyrx}
             f(u)^{r}_{x}=W_{C,y,x}F_{C,x,y}(f(u))=r_{u,v}f(u)_{x}.\end{equation}  Here $f(u)_{x}$ is the same complex number value in $\bar{C}_{x}$ as $f(u)$ is in $\bar{C}_{y}.$

             It follows that the global integral of $f$ over  $\{\phi(y):y\epsilon M\}$,  referred to $\bigcup_{x},$ is given by \begin{equation}\label{IfGx}I_{f,\theta,x}=\int_{x}(e^{\theta_{x}(u_{x})-\theta_{x}(v)})f(u_{x})du_{x}.\end{equation} This integral is over $\phi_{x}(M)$  or $\bar{R}^{4}_{x}.$ The integration variable, $u_{x},$ ranges over all coordinate values in $\phi_{x}(M).$ It is related to $u$ in $\phi_{y}(M)$ by $u_{x}=\phi_{x}(\phi_{y}^{-1}(u)).$

             The scalar field, $\theta_{x}$ has domain $\phi_{x}(M)$ and range $\bar{R}_{x}$. It is equivalent to $\theta$ in that $\theta(u)$ is the same real number value in $\bar{R}_{y}$ as $\theta_{x}(\phi_{x}(\phi_{y}^{-1}(u)))$ is in $\bar{R}_{x}.$  Here, as before, $\phi(y)=u$ is a coordinate value in $\phi_{y}(M).$

             In order to avoid confusion about the effect of scaling, it is necessary to be very explicit in the representation of the integrand in Eq. \ref{IfGx}. The steps in obtaining the integral are given by \begin{equation}\label{IntfGx}\begin{array}{l}I_{f,\theta,x}=\int W_{C,y,x}F_{C,x,y} (f(\phi(y))\times_{y}d\phi(y))=\int_{x} W_{C,y,x}(f(u_{x})_{x}\times_{x}du_{x})\\\\\hspace{1cm}=\int_{x} (r_{u_{x},v}f(u_{x})_{x})\frac{\times_{x}} {r_{u_{x},v}}(r_{u_{x},v}du_{x})=\int_{x}r_{u_{x},v}f(u_{x})_{x}du_{x}.\end{array}\end{equation}

             Note that  the whole term in the integrand must be scaled.  One cannot scale the factors separately and then multiply them afterwards.  The scaling of the multiplication operation must also be included. If this is done, then the correct number, one, of scaling factors is obtained, as shown in Eq. \ref{IfGx}.

             The same effect of scaling shows up in the mapping of the global expression for a wave packet $\psi =\int_{M}|y\rangle dy$ to an equivalent expression of an integral over $\phi(M)$ as $\psi=\int_{\phi}|\phi(y)\rangle d\phi_{y}.$ The equivalent local expression  as an integral over $\phi_{x}(M)$ is given by Eqs. \ref{psix} or \ref{FIhxphi}.

             Inclusion of scaling replaces these expressions for $\psi_{x}$ by \begin{equation}\label{psiScx}
             \psi^{\theta}_{x}=\int_{x}(e^{\theta_{x}(u_{x})-\theta_{x}(v)})|u_{x}\rangle_{x}\psi_{x}(u_{x})du_{x}.\end{equation} The presence of the scaling factor follows from the effects of scaling on Hilbert spaces shown in Eq. \ref{bHr}.

             These two examples  are representative of the effects of scaling on theoretical descriptions of  physical quantities.  Other examples, including some effects on geometry, are given in \cite{BenINTECH,BenNOVA}. As noted in the introduction, this work differs from the earlier work in that special emphasis is placed on the role played by coordinate charts.  These lift global expressions of nonlocal quantities defined over $M$ to local expressions defined over $\phi_{x}(M)$ in a universe, $\bigcup_{x}$ at a location, $x$ on $M$.

             There  is one effect of scaling on integral expressions that remains to be discussed.  In the description of  the effects of scaling the differential line element $d\phi(y)$ is simply replaced by $du_{x}.$ One might think  that, because $d\phi(y)$ is itself nonlocal, an extra scaling factor is needed to account for this nonlocality.

             This is not the case.  From Eq. \ref{dphiy} one sees that $$d\phi(y)=\phi(y+dy)-\phi_{y}.$$  The corresponding representation in $\phi_{x}(M)$ is given by Eq. \ref{dphiy} as $$d\phi_{x}(y)=\phi_{x}(y+dy)-\phi_{x}(y).$$ Inclusion of scaling replaces this expression by \begin{equation}\label{dphixyux}d\phi_{x}^{r}(y)\equiv du_{x}^{r}= e^{\theta_{x}(u_{x}+du_{x})-\theta_{x}(u_{x})} (u_{x}+du_{x})-u_{x}.\end{equation}

             Replacement of the scaling exponent by $(d\theta(u_{x})/du_{x})du_{x}$ and expansion of the scaling factor in powers of the exponent gives \begin{equation}\label{duxrt}du_{x}^{r}=(1+(d\theta(u_{x})/du_{x})du_{x}+\cdots)(u_{x}+du_{x})-u_{x}=du_{x}.
             \end{equation} Terms containing $du_{x}$ to the first and higher powers are neglected as  they are vanishingly small.  This shows that scaling inside $d\phi(y)$ has no effect and can be ignored.

             \section{Discussion}\label{D}
             There are many aspects of local mathematics and number scaling that need more work.  For example one would like to know what physical property, if any, is described by the scalar field, $\theta$.  As was noted, there are many different scalar fields proposed in physics. These include the Higg's boson, quintessence, the inflaton, etc. \cite{Higgs,PLorenz,Linde}. Whether any of these fields can be described by $\theta$ remains to be seen.

             It is important to keep in mind the fact that all descriptions of physical systems at far away cosmological points are described by us, as observers in a local reference region, We use the mathematics in $\bigcup_{x}$ where $x$ is any point in the region. The magnitudes of physical quantities at far away locations are represented by us, locally, as scaled magnitudes.  For example, a far away distance element $ds^{2}_{y}$ at space and time $y=\mathbf{y},t$ has a scaled representation  at $x=\mathbf{x},t'$  given by
             \begin{equation}\label{ds2y}(ds^{2}_{y})_{x}^{\theta}=e^{\theta(\mathbf{y},t)_{x}-\theta(\mathbf{x},t')}ds^{2}_{x}.\end{equation} Here $\mathbf{x}$ is any space point in our local region and $t'$ is the present age of the universe, about $14\times 10^{9}$ years.

             From this equation one sees that the space and time dependence of $\theta$ can be chosen so that for times, $t$ right after the big bang, the local $\mathbf{x},t'$  representation of  time $t$ distance elements, includes a  large contraction factor.  As an example let $\theta(\mathbf{x},t)\rightarrow -\infty$ as $t\rightarrow 0,$ the time of the big bang. One can mimic inflation by letting $\theta(\mathbf{x},t)$  increase at a very fast rate, even exponentially fast.  This results in our representing distance elements as increasing very fast as time increases from  the big bang. The rate of increase of $\theta$ can then slow down to show a slow and continuing increase of distance elements at the present time.

             In general, the space and time dependence of $\theta$ can be used to describe the functional  dependence on space and time locations of the magnitudes of  many physical quantities when referred to  a local coordinate system, $\phi_{x}(M).$ The functional dependence may be depend on the physical quantity being considered. However, the underlying manifold, $M$ remains unchanged.

             So far general relativity has not been treated here.  This needs to be changed.  Expansion of this work to include general relativity is work for the future. Also quantum mechanical effects need to be more deeply embedded into the setup than is shown by the wave packet example described earlier.

             A fact that may help in expanding the description of local mathematics and scaling is that the inclusion of number scaling into gauge theories results  in a vector field, $\vec{A}(x),$   description of scaling. The scalar field $\theta$ was introduced by the additional assumption  that $\vec{A}(x)$ is conservative and is the gradient of a scalar field, $\theta$ \cite{BenNOVA}. If one drops this assumption, then the description of space time integrals gets much more complicated because space and time integrals must include path dependence, perhaps as path integrals.

             Another somewhat more philosophic point concerns the relation of observers to the manifold, $M$ and the local universes.  As shown in Fig. \ref{SPIE41},  observers are at different locations, $x,y$ on $M$ in the cosmologically local reference region.  The mathematics available to an observer at $x$ is that in $\bigcup_{x}.$  Here $M$ is supposed to represent the actual physical cosmological space and time with observers at positions in the local reference region.

             As long as observers  in the local region are describing properties of physical systems within the universe, this picture is valid. The possibility of problems arises when observers attempt to describe the evolution and dynamics of the universe as a whole.This requires that mathematical descriptions of $M$, and all systems within the universe, including the observers, and a description of the universe itself, be described within a local universe.

             Achieving this seems problematic at least from a mathematical viewpoint.  There is no problem in including observers as physical systems.  However an observers description of the physical universe is a metamathematical description of a theory describing the universe.  As such it is outside the universe, and it places some aspects of the observer outside the universe. One aspect of this is that it includes  lifting $M$ past an observer into a local universe so that mathematically the observer is now external to
             $M$.

             This problem might be solved if one can find a mathematical description or model, in $\bigcup_{x},$ of an observer, at $x$ and outside of  $\bigcup_{x},$  describing a mathematical model of the  physical universe in $\bigcup_{x}$. This seems problematic at best.

             \section{Conclusion}
             In this work some consequences of number scaling and the local availability of mathematics were discussed. The gauge theory origins of these assumptions were summarized.  Emphasis was placed on the use of coordinate charts to lift global descriptions on a background space and time manifold to descriptions based on coordinate systems within local mathematical universes. The role of observers in giving meaning to mathematics locally was noted. Some of the effects of number scaling described by a scalar field were described.  The field $\theta,$ appears in gauge theories as a scalar boson with spin $0$.

             The affect of $\theta$ on descriptions of space and or time integrals or derivatives in physics, and the lack of experimental evidence for $\theta,$ place restrictions on the field.  The coupling constant of $\theta$ with other fields in gauge theories must be very small compared to the fine structure constant.  Also the gradient of $\theta$ must be very small in a local region of cosmological space and time that includes us as observers.  So far, there are no restrictions on $\theta$ at locations far away from our local region.

             Additional items  and open problems were described in the discussion section.  As they show, there is much work needed to further develop consequences of local mathematics and number scaling.  It is hoped to pursue these in the future.

            \section*{Acknowledgement}
            This work was supported by the U.S. Department of Energy,
            Office of Nuclear Physics, under Contract No. DE-AC02-06CH11357.

            \end{document}